\documentclass[sn-mathphys-ay]{sn-jnl}

\usepackage[abs]{overpic}
\usepackage{tabularx}
\usepackage{graphicx}%
\usepackage{multirow}%
\usepackage{amsmath,amssymb,amsfonts}%
\usepackage{amsthm}%
\usepackage{mathrsfs}%
\usepackage[title]{appendix}%
\usepackage{xcolor}%
\usepackage{textcomp}%
\usepackage{manyfoot}%
\usepackage{booktabs}%
\usepackage{algorithm}%
\usepackage{algorithmicx}%
\usepackage{algpseudocode}%
\usepackage{listings}%
\usepackage{scalerel}
\usepackage{tikz}
\usetikzlibrary{svg.path}
\usepackage{tikz}
\usetikzlibrary{calc} 

\usetikzlibrary{svg.path}

\definecolor{orcidlogocol}{HTML}{A6CE39}
\tikzset{
	orcidlogo/.pic={
		\fill[orcidlogocol] svg{M256,128c0,70.7-57.3,128-128,128C57.3,256,0,198.7,0,128C0,57.3,57.3,0,128,0C198.7,0,256,57.3,256,128z};
		\fill[white] svg{M86.3,186.2H70.9V79.1h15.4v48.4V186.2z}
		svg{M108.9,79.1h41.6c39.6,0,57,28.3,57,53.6c0,27.5-21.5,53.6-56.8,53.6h-41.8V79.1z M124.3,172.4h24.5c34.9,0,42.9-26.5,42.9-39.7c0-21.5-13.7-39.7-43.7-39.7h-23.7V172.4z}
		svg{M88.7,56.8c0,5.5-4.5,10.1-10.1,10.1c-5.6,0-10.1-4.6-10.1-10.1c0-5.6,4.5-10.1,10.1-10.1C84.2,46.7,88.7,51.3,88.7,56.8z};
	}
}

\newcommand\orcidicon[1]{\href{https://orcid.org/#1}{\mbox{\scalerel*{
				\begin{tikzpicture}[yscale=-1,transform shape]
					\pic{orcidlogo};
				\end{tikzpicture}
			}{|}}}}

\usepackage{hyperref}
\usepackage{natbib}
\usepackage{svg}
\setcitestyle{authoryear,round}

\begin{document}

\title[Article Title]{Full-domain POD modes from PIV asynchronous patches}


\author*[1]{\fnm{Iacopo} \sur{Tirelli}}\email{iacopo.tirelli@uc3m.es \orcidicon{0000-0001-7623-1161}}

\author[2]{\fnm{Adrian} \sur{Grille Guerra}}\email{a.grilleguerra@tudelft.nl  \orcidicon{0009-0006-1136-8988}}

\author[1]{\fnm{Andrea} \sur{Ianiro}}\email{aianiro@ing.uc3m.es \orcidicon{0000-0001-7342-4814}}
\author[2]{\fnm{Andrea} \sur{Sciacchitano}}\email{a.sciacchitano@tudelft.nl   \orcidicon{0000-0003-4627-3787}}

\author[2]{\fnm{Fulvio} \sur{Scarano}}\email{f.scarano@tudelft.nl    \orcidicon{0000-0003-2755-6669}}

\author*[1]{\fnm{Stefano} \sur{Discetti}}\email{sdiscett@ing.uc3m.es \orcidicon{0000-0001-9025-1505}}

\affil*[1]{\orgdiv{Department of Aerospace Engineering}, \orgname{Universidad Carlos III de Madrid}, \orgaddress{\street{Avda. Universidad 30}, \city{Leganés}, \postcode{28911}, \state{Madrid}, \country{Spain}}}

\affil[2]{\orgdiv{Faculty of Aerospace Engineering}, \orgname{Delft University of Technology}, \city{Delft}, \country{the Netherlands}}


\abstract{
A method is proposed to obtain full-domain spatial modes based on Proper Orthogonal Decomposition (POD) of Particle Image Velocimetry (PIV) measurements performed at different (overlapping) spatial locations. This situation occurs when large domains are covered by multiple non-simultaneous measurements and yet the large-scale flow field organization is to be captured. The proposed methodology leverages the definition of POD spatial modes as eigenvectors of the spatial correlation matrix, where local measurements, even when not obtained simultaneously, provide each a portion of the latter, which is then analyzed to synthesize the full-domain spatial modes. The measurement domain coverage is found to require regions  overlapping by \textbf{$50-75\%$} to yield a smooth distribution of the modes. The procedure identifies structures twice as large as each measurement patch. The technique, referred to as \textit{Patch POD}, is applied to planar PIV data of a submerged jet flow where the effect of patching is simulated by splitting the original PIV data. Patch POD is then extended to $3D$ robotic measurement around a wall-mounted cube. The results show that the patching technique enables global modal analysis over a domain covered with a multitude of non-simultaneous measurements.}


\maketitle

\section{Introduction}
\label{Sec:Intro}

Characterizing turbulent flows poses a significant challenge due to the wide range of spatial and temporal scales involved, which expand with increasing Reynolds number. In Particle Image Velocimetry \citep[PIV,][]{raffel2018particle}, the range of measured scales is constrained by hardware and physical limitations. These include, among others, the sensor size and the mean particle spacing on the images. The ratio between the largest and smallest measurable scales is referred to as the Dynamic Spatial Range (DSR), as noted by \citet{adrian1997dynamic}. The smallest measurable scale depends on the particle concentration and its ability to sample the flow field, while the largest can be extended by using larger camera sensors and/or increasing the field of view by reducing optical magnification. It can be shown that the turbulent Reynolds number for which complete resolution of scales can be achieved scales linearly with the product of the DSR and the Dynamic Velocity Range \citep{westerweel2013particle}, whereby the latter is defined as the ratio between maximum and minimum measurable velocities \citep{adrian1997dynamic}. 

An approach to enhance the spatial resolution (DSR) involves stitching together multiple views of the flow domain. This method uses several cameras to simultaneously capture different regions, which are then combined to create a complete picture of the flow field \citep{cardesa20122d,ferreira2020piv,li2021physics}. However, to capture these regions simultaneously, the entire flow field must be illuminated, and optical access for all cameras must be ensured. These requirements, along with the significant setup costs, represent the main drawbacks of this approach. For this reason, when the goal of the measurement is primarily to extract flow statistics, asynchronous (i.e. performed at different times) acquisitions are often preferred.

A simple implementation of this principle is based on traversing a Stereoscopic PIV (SPIV) system that includes the light sheet and the cameras, such to capture different slices of the flow field \citep{cardano2008piv,ostermann2016time,sellappan2018time,zigunov2022hysteretic}. At each slice,  a sequence of recordings is acquired, with a sample size large enough to ensure statistical convergence. The measurements are carried out at various cross-plane coordinates, allowing for the acquisition of multiple $2D-3C$ (two-dimensional three-components) velocity fields. The statistics of the fields are then stitched together to elaborate flow statistics on the full-covered domain. More recent studies \citep{rousseau2020scanning,zigunov2023continuously} show that it is possible to obtain an average $3D-3C$ volumetric flow field by applying spatiotemporal averaging to the stereoscopic PIV fields recorded while a traverse translates the laser plane continuously scanning the volume, rather than stopping at each plane to achieve a converged average.

Three-dimensional PIV and Lagrangian Particle Tracking (LPT) are increasingly used for the investigation of complex and unsteady three-dimensional flow fields \citep{discetti2018volumetric,schroder20233d}. Recent advances in volumetric PIV using helium-filled soap bubbles (HFSB) as flow tracers \citep{scarano2015use} enable to afford larger measurement domains up to the order of $1 m^3$. Nonetheless, measurements in domains with complex geometry and limited optical access are still elusive due to the lengthy procedures associated with system setup and calibration. The recently introduced coaxial volumetric velocimetry \citep[CVV,][]{schneiders2018coaxial} is well suited for robotic manipulation of the velocimeter that aims to circumvent the problem of optical access by performing several, local, volumetric measurements without the need for recalibration after the repositioning of the probe. This approach has been demonstrated in the work of \citet{jux2018robotic}, where the flow around a full-scale replica of a professional cyclist was surveyed. On the downside, robotic PIV performs independent measurements at each position of the probe, resulting into separated, often partly overlapping, patches (or sub-domains). As previously discussed, flow statistics are obtained by “stitching” measurements performed in independent sub-domains, assuming ergodicity and measuring at different time instants (asynchronous) in different regions the probability density functions of the velocity fluctuations. 

However, the partition of the investigated volume in patches prevents the visualization of any global, instantaneous flow feature that exceeds the size of an individual measurement region. 
This limitation has hindered, so far, the interpretation of coherent flow structures spanning more than a single measurement region, or \textit{patch}, using, for instance,  the Proper Orthogonal Decomposition \citep[POD,][]{lumley1967structure}.  

In the PIV community, this technique has gained widespread popularity due to an efficient algorithm for computing the POD, known as the ``snapshot method" \citep{sirovich1987turbulence}. 

For cases in which the domain is covered with non-simultaneous measurements, however, the straightforward application of the snapshot POD is not feasible. A snapshot matrix can still be built by considering as a common grid the set of points onto which the individual patches are stitched together. This matrix is massively gappy, with large blocks of missing data that hinder an accurate decomposition. 

Among the variants of POD, Gappy POD \citep{everson1995karhunen, venturi2004gappy} is notable for its ability to handle incomplete data by iteratively filling in gaps. The main idea is that the snapshots can be restored in an iterative process by performing POD with incomplete data, and low-order reconstruction with a progressively larger number of modes. However, its direct application in this context proves ineffective, as the only data available are highly localized within the patch domain. Similar considerations apply to other algorithms with similar rationale \citep{raben2012adaptive,cortina2021sparse}.

In this work, we propose to estimate spatial POD modes directly decomposing the two-point correlation matrix. Following the intuition of \citet{lumley1981coherent}, citing the review from \citet{george201750}, ``The actual determination of the eigenfunctions needs only statistics, which in turn requires only simultaneous measurements at two space-time points''. In principle, spatial POD modes can be obtained simply by scanning a domain with two hot-wires capturing simultaneously. This principle was leveraged by \citet{stokes1999multi} in a backward-facing-step experiment, and it required $33$ days of tests to cover a grid of $13 \times 750 \times 17$ measurement points. Robotic PIV allows covering a much larger number of simultaneous points, thus we expect this same principle can apply to volumetric measurements performed over patches while scanning a larger flow volume. We will investigate the conditions that make the implementation of such method viable, with particular attention to the size of patches and their mutual overlap. The key enabler is the progressive discovery of portions of the two-point correlation matrix. We refer to this method as ``Patch POD''.

The method is illustrated in \S\ref{sec:method}. 
In \S\ref{Sec:validation} we assess the proposed methodology employing a dataset of planar PIV measurements of a submerged jet. In this case the measurement domain is subdivided into patches, in order to simulate the case of asynchronous measurements in regions to be stitched together. In \S\ref{3D} the algorithm is applied to three-dimensional measurements performed using robotic PIV on the flow around a wall-mounted cube.

\section{Methodology} \label{sec:method}
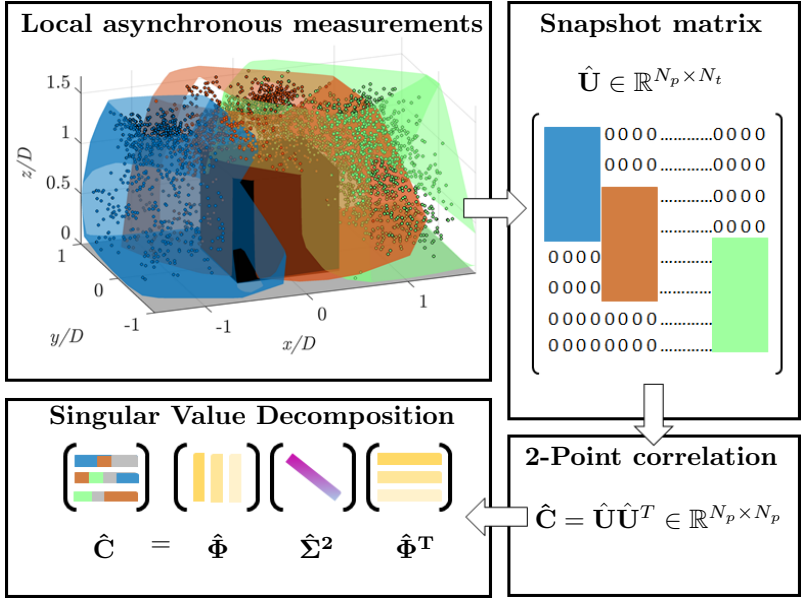
\begin{figure}[t]
    \centering
    \begin{overpic}[width=0.8\textwidth, unit = 1mm]{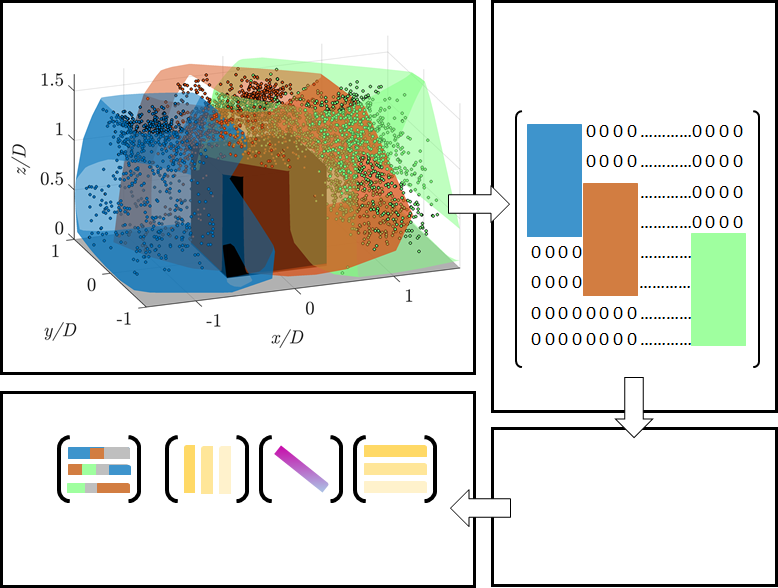}
    \put(0,75){\parbox{65mm}{\centering \textbf{{Local asynchronous measurements}}}} 
    \put(0,23){\parbox{65mm}{\centering \textbf{{Singular Value Decomposition}}}} 
    \put(68,75){\parbox{34mm}{\centering \textbf{{Snapshot matrix}}}} 
    \put(68,68){\parbox{34mm}{\centering {{$\hat{\mathbf{U}}\in \mathbb{{R}}^{N_p\times N_t}$}}}} 
    \put(68,18){\parbox{34mm}{\centering \textbf{{2-Point correlation}}}} 
    \put(69,10){\parbox{34mm}{\centering {{$\mathbf{\hat{C}} = \hat{\mathbf{U}}\hat{\mathbf{U}}^T\in \mathbb{R}^{N_p\times N_p}$}}}} 
    \put(8,6){\parbox{10mm}{\centering {{$\mathbf{\hat{C}}$}}}} 
    \put(18,6){\parbox{5mm}{\centering {{=}}}}
    \put(23,6){\parbox{10mm}{\centering {{$\mathbf{\hat{\Phi}}$}}}} 
    \put(36,6){\parbox{10mm}{\centering {{$\mathbf{\hat{\Sigma}^2}$}}}}
    \put(49,6){\parbox{10mm}{\centering {{$\mathbf{\hat{\Phi}^T}$}}}} 
    \end{overpic}
    \caption{Flowchart of the proposed algorithm. First step:  measuring the velocity fields from local asynchronous regions, i.e., patches; second step: building the snapshot matrix  $\hat{\mathbf{U}} \in \mathbb{R}^{N_p \times N_t}$, representing the entire DOI, with zeros filling areas outside the patches; 
    third step: compute two-point correlation; fourth step decomposition trough SVD to extract full-domain POD spatial modes.}
    \label{fig:flowchart}
\end{figure}

The principle of the proposed methodology is graphically exemplified in the flowchart shown in Fig.~\ref{fig:flowchart}. We assume that instantaneous snapshots of the velocity vector field are obtained asynchronously in different regions of the domain of interest (DOI). The velocity vector may be either computed by spatial cross-correlation or tracking of individual particles, which are later mapped onto Eulerian grids by averaging the vectors that, at a given time instant, fall within a specific bin (binning). For simplicity, it is assumed that the velocity vectors are available on a global Cartesian grid: $\mathbf{x} = (x, y) \in \mathbb{R}^{n_x \times n_y}$ in $2D$ or $\mathbf{x} = (x, y, z) \in \mathbb{R}^{n_x \times n_y \times n_z}$ for 3D measurements, where $n_x$, $n_y$ and $n_z$ are the number of grid points along the corresponding directions $x$, $y$ and $z$. The measured dataset is reorganized into a matrix $\mathbf{U}$ of dimensions $N_p$ (total number of grid points) and $N_t$ (total number of snapshots). For the case of domains covered with asynchronous measurements in patches, each snapshot will contain information only in the set of points corresponding to the patch for that time instant. 

Subsequently, the snapshot matrix is composed of the original velocity fields masked by zeros in the region outside of the patch. This can be written as a Hadamard product between $\mathbf{U}$ and a Dirac delta function matrix $\delta$, with zeros entries in the locations without velocity information:
\begin{equation}
\mathbf{\hat{U}} = \delta \odot \mathbf{U},
\end{equation}
where $\odot$ denotes the Hadamard product and $\mathbf{\hat{U}} \in \mathbb{R}^{N_p \times N_t}$ represents the masked version of $\mathbf{U}$. 

To compute spatial POD modes one could compute the eigenvector of the spatial correlation matrix $\mathbf{C}=1/N_t\cdot\mathbf{U} \mathbf{U}^\top$ $\in \mathbb{R}^{N_p \times N_p}$ which contains the correlations between the velocities at the different points of the domain. While the matrix $\mathbf{U}$ is not available, $\mathbf{C}$ can be approximated employing $\mathbf{\hat{U} \hat{U}^\top}$. This approximation of the spatial correlation matrix $\mathbf{\hat{C}} \in \mathbb{R}^{N_p \times N_p}$ would generate entries that are inconsistent with those of $\mathbf{U} \mathbf{U}^\top$ since each element of $\mathbf{\hat{C}}$ is actually evaluated only on a portion of the $N_t$ snapshots. To consider this, each non-zero element of the correlation matrix $\mathbf{\hat{U}}\mathbf{\hat{U}}^\top$ can be divided by the corresponding number of occurrences of non-zero entries. This rescaling is similar to the one leveraged by \cite{cortina2021sparse}. 
Non-zero entries of $\mathbf{\hat{C}}$ are limited to pairs of points belonging to the same patch or to points belonging to overlapping patches; pairs of points belonging to non-overlapping patches will be assigned a correlation value equal to zero, as in the following equation: 

\begin{equation}
\mathbf{\hat{C}} = \left[\left( \mathbf{\hat{U}}\mathbf{\hat{U}}^\top\right) \oslash \mathbf{N_{occ}} \right] \odot \mathbf{\delta_{ent}}.
\label{eq:2Pcorr}
\end{equation}
$\oslash$ is the Hadamard division operator, $\mathbf{N_{occ}} \in \mathbb{R}^{N_p \times N_p}$ is the matrix containing in each element $i,j$ the number of occurrences of non-zero entries in the corresponding elements $i$ and $j$ of  $\mathbf{\hat{U}}\mathbf{\hat{U}}^T$, and $\mathbf{\delta_{ent}}$ is the matrix containing ones wherever $\mathbf{N_{occ}}(i,j)>0$ and zeros elsewhere. 

Approximated spatial modes are then computed as eigenvectors of $\mathbf{\hat{C}}$:
\begin{equation}
\mathbf{\hat{C}} = \mathbf{\hat{\Phi} \hat{\Sigma}}^2 \mathbf{\hat{\Phi}}^\top,
\label{eq:2Pcorr}
\end{equation}
For the latter, the symbol $\hat{\cdot}$ is used to refer  to not complete field. In Eq. \eqref{eq:2Pcorr} $\mathbf{\hat{\Phi}} \in \mathbb{R}^{N_p \times N_p}$ is the matrix whose columns are the spatial modes $\hat{\phi}_i$, and $\mathbf{\hat{\Sigma}} \in \mathbb{R}^{N_p \times N_p}$ the diagonal matrix whose elements $\hat{\sigma}_i$ represent the singular values corresponding to spatial modes.

It must be remarked that this approach allows solely the recovery of the global spatial modes. The estimation of the POD temporal modes can only be performed within each patch individually and not globally.

\section{Assessment by experiments}
\label{Sec:validation}
\subsection{Free jet}\label{SubSec:jet}

\begin{figure}[t]
    \centering
    \fontsize{9}{10}
    \includesvg[width=0.7\linewidth]{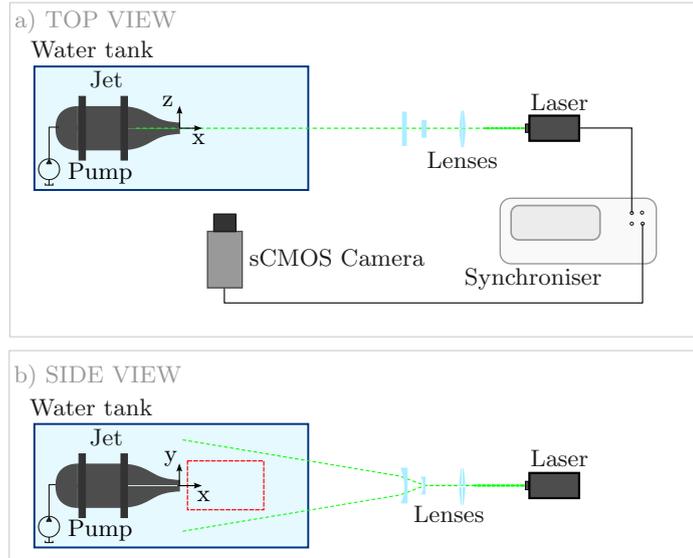}
    \caption{Submerged water jet-flow system. Sketch of the experimental set-up: in red the common region onto which the obtained velocity fields are interpolated. Figure adapted from \cite{franceschelli2024assessment}.}
    \label{fig:exp}
\end{figure}

\begin{table}[t]
\centering
\caption{Experimental setup details.}
\begin{tabularx}{\textwidth}{|c|X|}
\hline
\textbf{Seeding} & Polyamide particles, $d = 56\,\mu$\text{m}. \\ \hline
\textbf{Working fluid} & Water. \\ \hline
\textbf{Exit diameter} & Circular shape, $D = 20$ \text{mm}. \\ \hline
\textbf{Exit velocity} & $U_j = 0.13$ m/s. \\ \hline
\textbf{Reynolds number} & $Re_D = 2,600$. \\ \hline
\textbf{Illumination} & LaserTree LT-40W-AA ($5W$ power). \\ \hline
\textbf{Imaging} & Andor Zyla sCMOS $5.5$ MP ($2560 \times 2160$) pixel array, $6.5 \times 6.5$ $\mu$m pixel size, resolution $9.25$ pix/mm. Objective Tokina $50$ mm lens. $f_\#=8$. \\ \hline
\textbf{Processing} & POD background removal \citep{mendez2017pod}, iterative multi-grid/multi-pass algorithms \citep{willert1991digital,soria1996investigation}, image deformation \citep{scarano2001iterative}, B-spline interpolation \citep{astarita2005analysis,astarita2007analysis} implemented in PaIRS \citep{paolillo2024pairs}. \\ \hline
\end{tabularx}
\label{tab:jet}
\end{table}

The method is validated here by making use of planar PIV measurements of a submerged jet, installed in an $80 \times 60 \times 40 \, \text{cm}^3$ water tank. The experimental setup is illustrated in Fig.~\ref{fig:exp} and summarized in Tab.~\ref{tab:jet}. The jet nozzle has a circular shape with an exit diameter of $D=20$ \text{mm}. The bulk velocity is set to $U_j \approx 0.13\,\text{m/s}$, resulting in a Reynolds number $Re_D = 2,600$.

The water flow is seeded with polyamide particles (neutrally buoyant and of  $56$$\mu m$ diameter). Illumination is provided with continuous laser (LaserTree LT-40W-AA, $5W$ power). The laser is controlled by a pulse generator, which defines the duration of the pulses ($\delta t = 1$ ms)  and the time interval between consecutive laser pulses ($\Delta t = 10$ ms), corresponding to a frequency of $100$ Hz. Additionally, the pulse generator synchronizes the system by providing the acquisition trigger signal for the PIV camera.

Particle images are captured by an sCMOS camera (Andor Zyla, $5.5$ Mpx, pixel pitch  $6.5\mu m$) equipped with a $50mm$ Nikon objective. Recording is performed at a rate of $100$ frames/s. The field of view is set to $160 \times 56 mm^2$ ($8 \times 2.8 D^2$) with a resolution of $9.25$ pixels/mm. At the given flow speed and frame rate, the particle tracers move of approximately $1.3$ mm ($1/15$ of the jet diameter) among subsequent frames and the measurement is considered time-resolved. The whole measurement sequence encompasses a duration T = $600$ seconds ($400$ jet diameters) with $60,000$ frames. The measurement domain is divided into patches and for each patch, $1500$ snapshots are randomly extracted from the sequence, removing any temporal coherence and resulting in a non-time-resolved subset. Each snapshot is only used once in the process, i.e. it is never observed by more than one patch. This approach yields independent sets of measurements composing the dataset. More details of the experimental dataset are reported in the work by \cite{franceschelli2024assessment}.

\begin{figure}
    \centering
    \begin{overpic}[scale = 1, unit = 1mm]{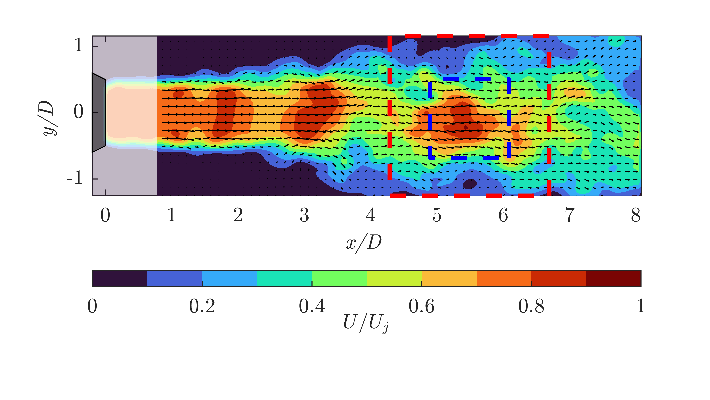}
    \end{overpic}
  
    \caption{Instantaneous streamwise velocity field contour and corresponding velocity vectors for the case of the submerged jet, normalized with the bulk velocity $U_j$. The blanked region near the jet exit, excluded from the DOI due to the presence of strong reflections, is included here to enhance visualization clarity. Examples of patches are illustrated: the red square represents a patch with $\mathcal{P}  = 2D$  while the blue one $\mathcal{P}  = D$.}
    \label{fig:mean_jet}
\end{figure}

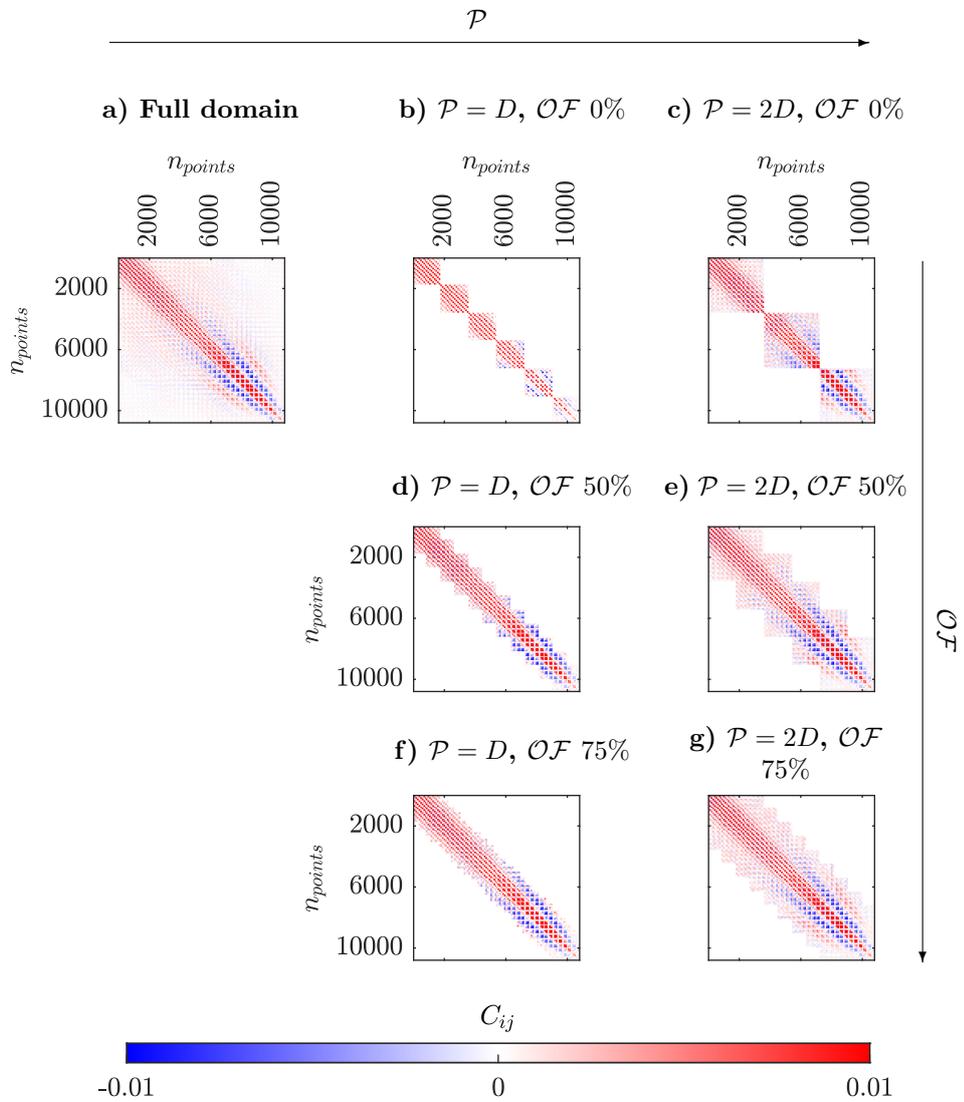
\begin{figure}[t]
    \centering
    \begin{overpic}[scale = 1, unit = 1mm]{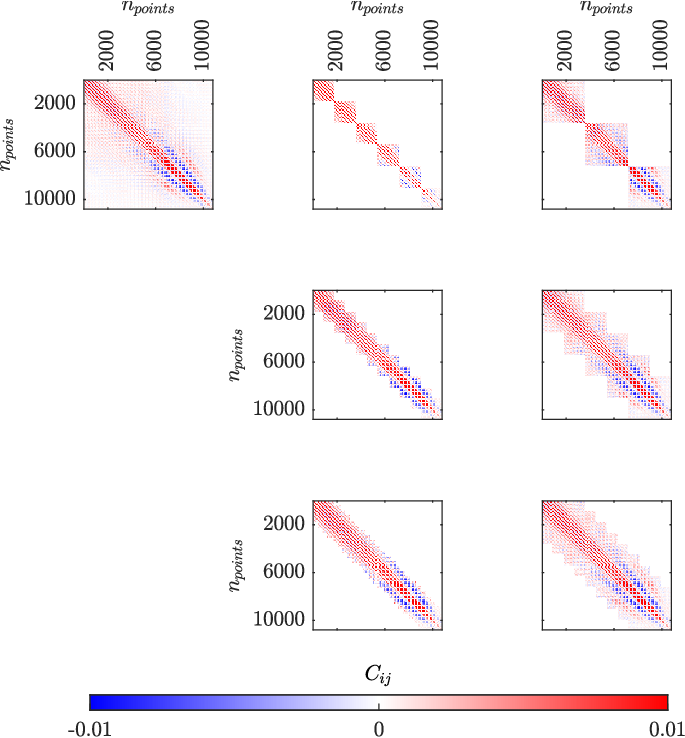}
     \put(5,130){\parbox{40mm}{\centering \textbf{{a) Full domain}}}}
    \put(50,130){\parbox{32mm}{\centering \textbf{{b) $\mathcal{P}=D$, $\mathcal{OF}$ $0\%$ }}}}
    \put(86,130){\parbox{32mm}{\centering \textbf{{c) $\mathcal{P}=2D$, $\mathcal{OF}$ $0\%$}}}}
    \put(50,80){\parbox{32mm}{\centering \textbf{{d) $\mathcal{P}=D$, $\mathcal{OF}$ $50\%$}}}}
    \put(86,80){\parbox{32mm}{\centering \textbf{{e) $\mathcal{P}=2D$, $\mathcal{OF}$ $50\%$}}}}
    \put(50,45){\parbox{32mm}{\centering \textbf{{f) $\mathcal{P}=D$, $\mathcal{OF}$ $75\%$}}}}
    \put(86,45){\parbox{32mm}{\centering \textbf{{g) $\mathcal{P}=2D$, $\mathcal{OF}$ $75\%$}}}}
    \put(13, 140){\vector(1, 0){100}}  
    \put(60, 142){\textbf{$\mathcal{P}$}}  
    \put(120, 111){\vector(0, -1){93}}  
    \put(122, 65){\rotatebox{-90}{\textbf{$\mathcal{OF}$}}}  
    \end{overpic}
   \caption{Spatial correlation matrices. The matrix obtained from the full domain (a) is compared against the patched case at $\mathcal{P} = D$, with overlap factors ranging from no overlap (b) to $50\%$ (d), and $75\%$ (f), as well as the analogous case at $\mathcal{P} = 2D$ (c), (e), and (g), respectively.}

    \label{fig:CorrelationJet}
\end{figure}

\subsection{Data analysis}
The use of a continuous wave laser in ``pulsed mode" may lead to slightly more elliptical particle shapes. This uncertainty affects both the reference and patched datasets equally, as they originate from the same measurements. While it might be expected slightly larger uncertainty in regions with higher flow velocity, this source of error is expected to be randomly distributed and thus affecting mostly high-order modes. Thus, it does not have any detectable impact on the analysis carried out in this section.
\label{subsec:assessment_jet}
\begin{figure}

    \centering
    \begin{tikzpicture}[scale = 2.3]
    \draw[->] (-1.8,0) -- (1.5,0) node[right] {$x/\mathcal{P}$}; 
    \draw[->] (-1.6,0.2) -- (-1.6,-1.5) node[right] {t};
   
    \foreach \x in {-1,-0.5,0,0.5,1} {
        \draw (\x,0.1) -- (\x,-0.1); 
        \node[below] at (\x,-0.1) {\x}; 
    }
        \foreach \x in {-0.75,-0.25,0.25,0.75} {
            \draw[thin] (\x,0.05) -- (\x,-0.05); 
        }
        
   \node at (-1.5,0.5) {a)};
    \shade[top color=black!20, bottom color=white] (-1.5,-0.30) rectangle (-0.5,-0.6);
    \draw[thick, black] (-1.5,-0.3) -- (-0.5,-0.3);
    \draw[thick, black] (-1.5,-0.3) -- (-1.5,-0.4); 
    \draw[thick, black] (-0.5,-0.3) -- (-0.5,-0.4); 

     \shade[top color=black!20, bottom color=white] 
        (-1,-0.5) rectangle (0,-0.8);
    \draw[thick, black] (-1,-0.5) -- (0,-0.5);
    \draw[thick, black](-1,-0.5) -- (-1,-0.6); 
    \draw[thick, black] (0,-0.5) -- (0,-0.6);

     \shade[top color=red!20, bottom color=white] 
        (-0.50,-0.7) rectangle (0.5,-1);
    \draw[thick, red] (-0.5,-0.7) -- (0.5,-0.7);
    \draw[thick, red] (-0.5,-0.7) -- (-0.5,-0.8); 
    \draw[thick, red] (0.5,-0.7) -- (0.5,-0.8);
    \draw[thick, teal] (-0,-0.7) -- (0.5,-0.7);

    \shade[top color=black!20, bottom color=white] 
        (1,-0.9) rectangle (0,-1.2);
    \draw[thick, black] (1,-0.9) -- (0,-0.9);
    \draw[thick, black] (1,-0.9) -- (1,-1); 
     \draw[thick, black] (0,-0.9) -- (0,-1); 
     \draw[thick, teal] (-0,-0.9) -- (0.5,-0.9);
       
    \shade[top color=black!20, bottom color=white] 
        (0.5,-1.1) rectangle (1.5,-1.4);
    \draw[thick, black] (0.5,-1.1) -- (1.5,-1.1);
    \draw[thick, black] (1.5,-1.1) -- (1.5,-1.2); 
    \draw[thick, black] (0.5,-1.1) -- (0.5,-1.2);

    \fill[blue] (0, -0.7) circle (0.03);
    \fill[blue] (0, -0.5) circle (0.03);
    \fill[blue] (0, -0.9) circle (0.03);
    \fill[blue,thick] (0.5, -0.7) circle (0.03);
    \fill[blue,thick] (0.5, -0.9) circle (0.03);
    \fill[blue,thick] (0.5, -1.1) circle (0.03);
    \fill[teal] (0.25, -0.7) circle (0.03);
    \fill[teal] (0.25, -0.9) circle (0.03);
    \draw[<->,dash dot, thick,blue](-1,-1.3) -- (1,-1.3)node[midway, below,blue] {$2\mathcal{P}$};
    \draw[dashed, blue] (-1,-1.3) -- (-1,-0.6); 
    \draw[dashed, blue] (1,-1.3) -- (1,-1); 
    \draw[<->,dash dot, thick,teal](-0.5,-1.5) -- (1,-1.5)node[midway, below,teal] {$1.5\mathcal{P}$};
    \draw[dashed, teal] (-0.5,-0.7) -- (-0.5,-1.5); 
    \draw[dashed, teal] (1,-1.5) -- (1,-1.3); 
\end{tikzpicture}

    \begin{tikzpicture}[scale = 2.3]
    \draw[->] (-1.8,0) -- (1.5,0) node[right] {$x/\mathcal{P}$}; 
    \draw[->] (-1.6,0.2) -- (-1.6,-1.5) node[right] {t};

    \foreach \x in {-1,-0.5,0,0.5,1} {
        \draw (\x,0.1) -- (\x,-0.1); 
        \node[below] at (\x,-0.1) {\x}; 
    }
        \foreach \x in {-0.75,-0.25,0.25,0.75} {
            \draw[thin] (\x,0.05) -- (\x,-0.05); 
        }
    \node at (-1.5,0.5) {b)};
    \shade[top color=black!20, bottom color=white] (-1,-0.30) rectangle (-0,-0.6);
    \draw[thick, black] (-1,-0.3) -- (-0,-0.3);
    \draw[thick, black] (-1,-0.3) -- (-1,-0.4); 
    \draw[thick, black] (-0,-0.3) -- (-0,-0.4); 

     \shade[top color=black!20, bottom color=white] 
        (-0.75,-0.5) rectangle (0.25,-0.8);
    \draw[thick, black] (-0.75,-0.5) -- (0.25,-0.5);
    \draw[thick, black] (-0.75,-0.5) -- (-0.75,-0.6); 
    \draw[thick, black] (0.25,-0.5) -- (0.25,-0.6);

     \shade[top color=red!20, bottom color=white] 
        (-0.50,-0.7) rectangle (0.5,-1);
    \draw[thick, red] (-0.5,-0.7) -- (0.5,-0.7);
    \draw[thick, red] (-0.5,-0.7) -- (-0.5,-0.8); 
    \draw[thick, red] (0.5,-0.7) -- (0.5,-0.8);
    \draw[thick, teal] (-0,-0.7) -- (0.5,-0.7);

    \shade[top color=black!20, bottom color=white] 
        (-0.25,-0.9) rectangle (0.75,-1.2);
    \draw[thick, black] (-0.25,-0.9) -- (0.75,-0.9);
    \draw[thick, black] (0.75,-0.9) -- (0.75,-1); 
     \draw[thick, black] (-0.25,-0.9) -- (-0.25,-1); 
     
    \shade[top color=black!20, bottom color=white] 
        (0,-1.1) rectangle (1,-1.4);
    \draw[thick, black] (0,-1.1) -- (1,-1.1);
    \draw[thick, black] (1,-1.1) -- (1,-1.2); 
    \draw[thick, black] (0,-1.1) -- (0,-1.2); 

    \shade[top color=black!20, bottom color=white] 
        (0.25,-1.3) rectangle (1.25,-1.6);
    \draw[thick, black] (0.25,-1.3) -- (1.25,-1.3);
    \draw[thick, black] (1.25,-1.3) -- (1.25,-1.4); 
    \draw[thick, black] (0.25,-1.3) -- (0.25,-1.4); 
     
    \shade[top color=black!20, bottom color=white] 
        (0.5,-1.5) rectangle (1.5,-1.8);
    \draw[thick, black] (0.5,-1.5) -- (1.5,-1.5);
    \draw[thick, black] (1.5,-1.5) -- (1.5,-1.6); 
    \draw[thick, black] (0.5,-1.5) -- (0.5,-1.6); 

    \fill[blue] (0, -0.3) circle (0.03);
    \fill[blue] (0, -0.5) circle (0.03);
    \fill[blue] (0, -0.7) circle (0.03);
    \fill[blue] (0, -0.9) circle (0.03);
    \fill[blue] (0, -1.1) circle (0.03);
    
    \fill[blue,thick] (0.25, -0.5) circle (0.03);
    \fill[blue,thick] (0.25, -0.7) circle (0.03);
    \fill[blue,thick] (0.25, -0.9) circle (0.03);
    \fill[blue,thick] (0.25, -1.1) circle (0.03);
    \fill[blue,thick] (0.25, -1.3) circle (0.03);

    \fill[blue,thick] (0.5, -0.7) circle (0.03);
    \fill[blue,thick] (0.5, -0.9) circle (0.03);
    \fill[blue,thick] (0.5, -1.1) circle (0.03);
    \fill[blue,thick] (0.5, -1.3) circle (0.03);
    \fill[blue,thick] (0.5, -1.5) circle (0.03);

    \fill[teal] (0.125, -0.5) circle (0.03);
    \fill[teal] (0.125, -0.7) circle (0.03);
    \fill[teal] (0.125, -0.9) circle (0.03);
    \fill[teal] (0.125, -1.1) circle (0.03);

    \fill[teal,thick](0.375, -0.7) circle (0.03);
    \fill[teal,thick] (0.375, -0.9) circle (0.03);
    \fill[teal,thick] (0.375, -1.1) circle (0.03);
    \fill[teal,thick] (0.375, -1.3) circle (0.03);

    \draw[<->,dash dot, thick,blue](-1,-1.6) -- (1,-1.6)node[midway, below,blue] {$2\mathcal{P}$};
    \draw[dashed, blue] (-1,-1.6) -- (-1,-0.4); 
    \draw[dashed, blue] (1,-1.6) -- (1,-1.2); 

    \draw[<->,dash dot, thick,teal](-0.75,-1.8) -- (1,-1.8)node[midway, below,teal] {$1.75\mathcal{P}$};
    \draw[dashed, teal] (-0.75,-0.6) -- (-0.75,-1.8); 
    \draw[dashed, teal] (1,-1.8) -- (1,-1.6); 
\end{tikzpicture}    
    \caption{Sketch of overlapped patches for the case $\mathcal{OF}$ = $50\%$ (a) and $\mathcal{OF}$ = $75\%$ (b). In red the reference, in black  the ones overlapped with it. Best case scenario (in blue): centre of the patch, able to extend the original patch by a factor $2\mathcal{P}$; worst case scenario (in teal): scales according to the overlap and adds to the original patch an $\mathcal{OF}$$\cdot\mathcal{P}$.}
\label{fig.overlap}
\end{figure}
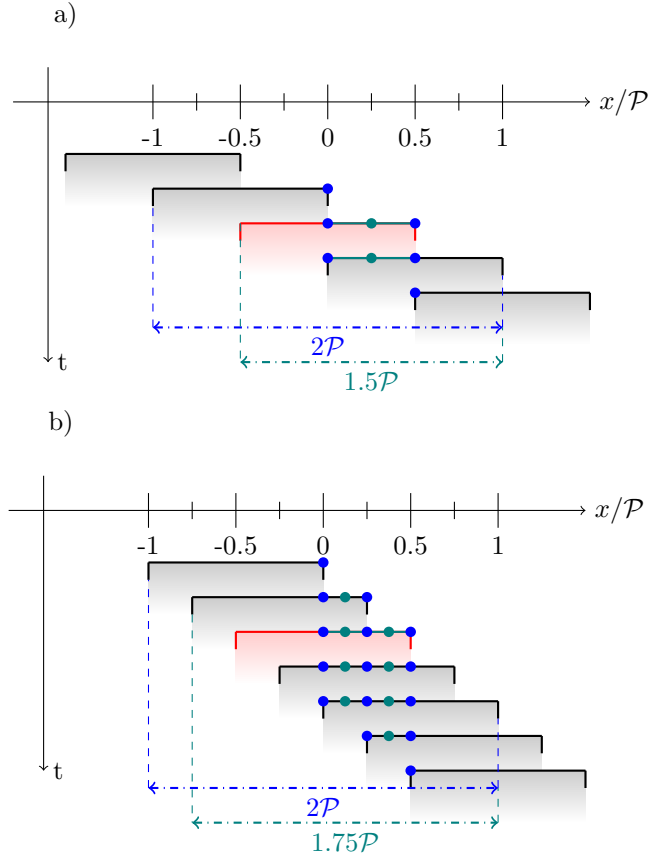

The velocity fields are obtained with cross-correlation with a final interrogation window of $32 \times 32$ pixels ($3.56 \times 3.56 mm^2$) and an overlap factor of $75\%$, resulting in a Cartesian description of the flow field on a grid of $60 \times 180$ velocity vectors with spacing of $0.9$ mm. With the latter resolution the jet diameter is covered by approximately $22$ data points. The patched measurement is simulated by partitioning the whole domain into square subdomains (i.e. patches). The patch dimension  $\mathcal{P}$ is indicated in diameters. Additionally, the overlap fraction $\mathcal{OF}$ (in percent) is monitored as a relevant parameter. 

Figure~\ref{fig:mean_jet} depicts a contour of an instantaneous streamwise velocity field and the corresponding velocity vectors within the full domain of acquisition: the blanked region near the jet exit is excluded from the DOI and the consequent analysis due to the presence of strong reflections. In addition, examples of patches are reported: $\mathcal{P} = D$  in blue and $\mathcal{P} = 2D$ in red.  These are analyzed using overlap factors of $\mathcal{OF} = [0\%,50\%,75\%$].


The structure of the correlation matrix $\hat{C}$ depends upon patch size and overlap factor. In addition, the number of entries depends upon the grid points that cover the patch. Patch size and overlap will determine the pattern of non-zero regions in the array.
Figure \ref{fig:CorrelationJet} shows the values of the spatial correlation coefficients $\hat{C_{ij}}$. The reference spatial correlation matrix, $\mathbf{C}$, obtained from the full domain measurement is presented in Fig.~\ref{fig:CorrelationJet}.a, while the results for the partitioned domain with $\mathcal{P} = D$ without overlap are shown in Fig.~\ref{fig:CorrelationJet}.b. Partitioning the domain allows to evaluate the correlations only within the individual patches. This results in computing only the regions of $\mathbf{C}$ near the diagonal. 

The overlap, however, extends this correlation up to $\pm0.5\mathcal{P}$ (Fig.~\ref{fig:CorrelationJet}.d,f). 
This effect is illustrated in Fig.~\ref{fig.overlap}.a with an exemplifying 1D sketch. Consider for instance the red patch. The best-case scenario (meaning that the correlation extends to the maximum distance, which in this case is $0.5\mathcal{P}$)  occurs for points located exactly in the center of the patch (blue circle). Indeed, when performing measurements in the red patch along, the two-point correlations between the center point and points located at $\pm 0.5\mathcal{P}$ are available. Furthermore, a $50\%$ overlap extends the effective patch size up to $2\mathcal{P}$ through the adjacent patches. A similar situation arises for points at the edges, also marked in blue, which in a $50\%$-overlap case are the centers of the adjacent patch. The worst case involves points located between the centers, for which portions of points available in the two-point correlation map cover only $1.5\mathcal{P}$. Among these, for the point positioned equidistantly between two patch centers (in green in the figure) a symmetric distribution of correlation coefficient is available ($\pm0.75\mathcal{P}$). For other intermediate points the same region is covered, but in general with asymmetric distribution.

Increasing the patch size to $2D$ while keeping the same overlap allows capturing a more complete picture of the correlation map, as shown in the last column in Fig.~\ref{fig:CorrelationJet}. However, increasing the overlap (up to $75\%$) yields the values around the diagonal with more homogeneity. In particular the minmax distance from the diagonal is reduced by increasing the overlap factor, as evident from Fig.~\ref{fig:CorrelationJet}.e-g. As detailed in Fig.~\ref{fig.overlap}.b, the maximum extension for the patch is still $2\mathcal{P}$, but the minimum correlation distance observed by each point is increased up to $1.75\mathcal{P}$ against the $1.5\mathcal{P}$ of the $50\%$ overlap case.

As a general rule, the size of the correlation region $\Tilde{\mathcal{P}}$ available for each point  can be computed as follows:

\begin{equation}
\begin{cases}
\Tilde{\mathcal{P}} = 2\mathcal{P} & \text{centers} \\
\Tilde{\mathcal{P}} = \mathcal{P} + \text{$\mathcal{OF}$} \cdot \mathcal{P} & \text{elsewhere}
\end{cases}
\end{equation}

This guideline is particularly useful for planning experiments, in which $\Tilde{\mathcal{P}}$ represents a good approximation of the largest lengthscale that can be properly observed in the POD modes. 

\subsection{Reference POD modes}

\begin{figure}[t]
    \centering
       \begin{overpic}[scale = 1, unit = 1mm]{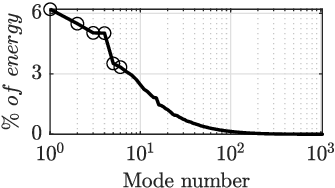}
    \put(-37.5,35){a)}
     
    \end{overpic}
 \vspace{0.3cm}   
    \begin{overpic}[scale = 1, unit = 1mm]{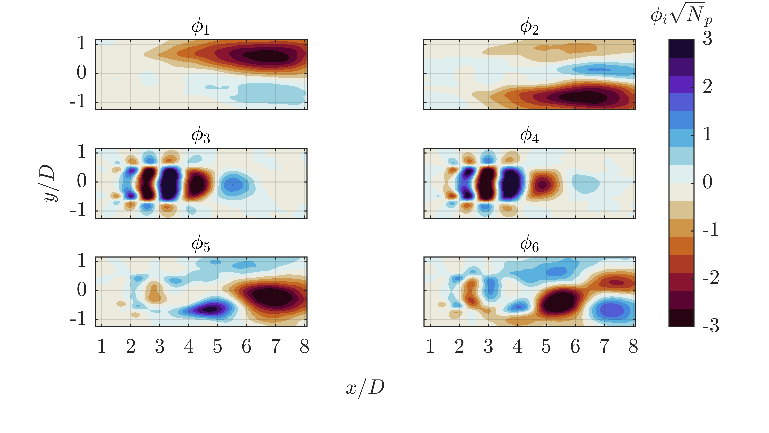}
    \put(0,70){b)}    
    \end{overpic}
    \caption{(a) The corresponding eigenvalues of $C$, normalized by their total sum, with circles indicating the contributions of the first six modes. (b) The first six spatial modes, $\phi_i$, derived from the decomposition of the reference spatial correlation matrix, $\mathbf{C}$.}

    \label{fig:ref_modes}
\end{figure}

 The data resulting from the  instantaneous measurement over the full domain is considered for reference, for which the spatial modes are extracted and physically examined. The decomposition of the spatial correlation matrix $\mathbf{C}$ provides the results shown in Fig.~\ref{fig:ref_modes}: the full-domain spatial modes $\phi_i$ (Fig.~\ref{fig:ref_modes}.b), normalized by their standard deviation ($\sqrt{N_p}$), and the corresponding eigenvalue distribution, expressed as a percentage of the total sum (Fig.~\ref{fig:ref_modes}.a). 

The first two modes accumulate approximately $11\%$ of the energy and they describe an observed flapping mode \citep{lynch2009pod}, which is a  dynamic instability characterized by large-scale oscillations in the jet flow. The modes feature elongated regions of momentum deficit or excess, which is interpreted as the cross-sectional view of a dynamical offset of the jet axis in flapping or circularly meandering motion \citep{batchelor1962analysis}.

Modes $3$ and $4$ accumulate $9\%$ of the energy and are coupled in the convective motion of the vortex rings forming in the free shear layer  \citep{violato2013three}. Such toroidal structures form near the nozzle exit, grow rapidly in the first diameter and travel along the axial direction of the jet maintaining symmetry for approximately three diameters as shown in Fig~\ref{fig:mean_jet}. The modes feature alternating values of positive velocity (excess) in the region inside the vortex ring and negative velocity (deficit) in the region between two rings. Furthermore, outside of the shear layer, the axial velocity exhibits a region of sign reversal ($\phi_3$ and $\phi_4$ in Fig.~\ref{fig:ref_modes}.b). The wavelength that separates two vortex rings is approximately $1D$, which agrees with the literature reference from \citet{schram2003aeroacoustics}. Finally, $\phi_3$ and $\phi_4$ are found in phase quadrature ($\pi/2$ shift) in accordance with the concept of convective motion.

These vortex rings are a distinctive feature of jet flows, particularly in the transitional regimes, and are sustained by the Kelvin-Helmholtz mechanism of shear layer instability. As these rollers travel downstream they tend to interact and enter a precessing motion in pairs (leapfrogging) as described in the work of \citet{schram2003aeroacoustics}, among others. During leapfrogging, the azimuthal modes grow rapidly (not captured in the present planar view), leading to the subsequent vortex filament distortion and breakup into more three-dimensional fluctuations. Vortex pairing is characterized by a double wavelength and a visible fluctuation of the jet core axis.

Modes $5$ and $6$ are interpreted as the result of vortex pairing and core breakdown  \citep{violato2013three}, which transfers energy to larger scales, and contributes to the growth and breakdown of these coherent structures. 

\subsubsection{Patch size and overlap }
\begin{figure}[t]
    \centering
    \begin{overpic}[scale = 1, unit = 1mm]{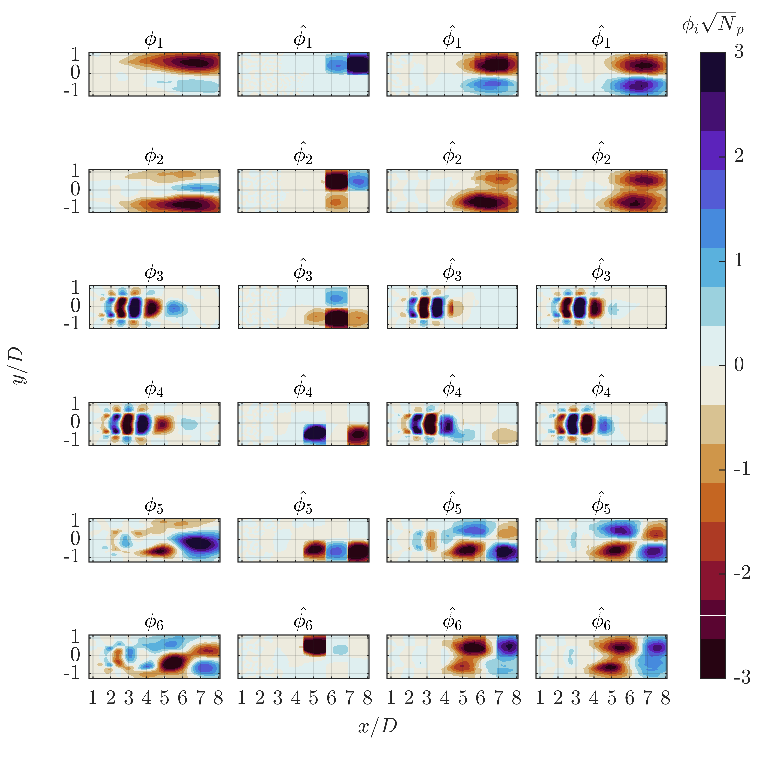}
    \put(13,130){\parbox{26mm}{\centering \textbf{{Full field}}}} 
    \put(38,130){\parbox{26mm}{\centering \textbf{{$\mathcal{OF}$ $0\%$}}}}
    \put(64,130){\parbox{26mm}{\centering \textbf{{$\mathcal{OF}$ $50\%$}}}}
    \put(88,130){\parbox{26mm}{\centering \textbf{{$\mathcal{OF}~75\%$}}}}
    \put(40,138){\parbox{70mm}{\centering \textbf{{$\mathcal{P}=D$}}}}
    \end{overpic}
    \caption{First six POD spatial modes of streamwise velocity component. Reference data (left column) is compared with patched POD ($\mathcal{P} = D$) for varying levels of overlap from no overlap ($2^\text{nd}$ column), to $50\%$ ($3^\text{rd}$ column) and $75\%$ ($4^\text{th}$ column). }
    \label{fig:phiJet_30}
\end{figure}

\begin{figure}[t]
    \centering
    \begin{overpic}[scale = 1, unit = 1mm]{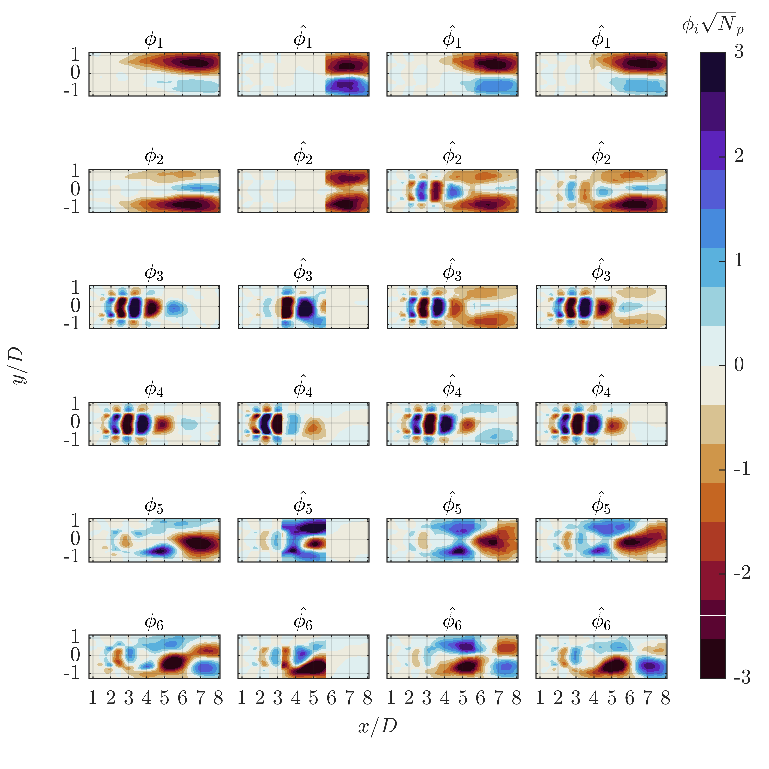}
    \put(13,130){\parbox{26mm}{\centering \textbf{{Full field}}}} 
    \put(38,130){\parbox{26mm}{\centering \textbf{{$\mathcal{OF}$ $0\%$}}}}
    \put(64,130){\parbox{26mm}{\centering \textbf{{$\mathcal{OF}$ $50\%$}}}}
    \put(88,130){\parbox{26mm}{\centering \textbf{{$\mathcal{OF}~75\%$}}}}
    \put(40,138){\parbox{70mm}{\centering \textbf{{$\mathcal{P}=2D$}}}}
    \end{overpic}
    \caption{First six POD spatial modes of streamwise velocity component. Reference data (left column) is compared with patched POD ($\mathcal{P} = 2D$) for varying levels of overlap from no overlap ($2^\text{nd}$ column), to $50\%$ ($3^\text{rd}$ column) and $75\%$ ($4^\text{th}$ column). }
    \label{fig:phiJet_60}
\end{figure}

\begin{figure}[t]
    \centering
  \begin{tikzpicture}
        \node[inner sep=0pt] (image) at (0,0) {\includegraphics[scale=1]{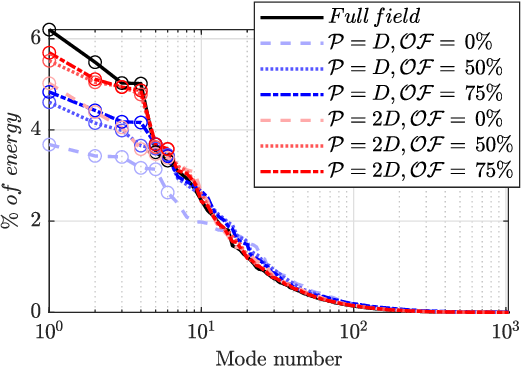}};
        
    \end{tikzpicture}
    
    \caption{Normalised eigenvalues spectrum. Reference data (black) $\mathcal{P} = D$ (blue) and $\mathcal{P} = 2D$ (red) are compared, with varying levels of overlap. Empty circles highlight the first $6$ modes. }
    \label{fig:eigjet}
\end{figure}

The methodology is evaluated through an analysis of the reconstructed spatial modes $\hat{\phi}$ (Figs.~\ref{fig:phiJet_30}-\ref{fig:phiJet_60}) and the associated eigenvalues (Fig.~\ref{fig:eigjet}), obtained by computing the square of the singular values $\hat{\sigma}$, normalized with the total sum of the reference case. These are the results of the POD decomposition of the correlation matrices $\mathbf{\hat{C}}$.

The spatial modes reconstructed with varying patch size and overlap are organized into two separate panels for each value of  $\mathcal{P}$. The reference modes $\phi_i$ are displayed in the first column, followed by columns with the reconstructed $\hat{\phi_i}$ for progressively increasing levels of $\mathcal{OF}$. The analysis involves the first six spatial modes.
 
 In the comparison proposed in Fig.~\ref{fig:phiJet_30}, the reconstruction with no overlap ($2^{nd}$ column) among patches produces modes that do not correspond to the reference distribution ($1^{st}$ column). This is ascribed to the discontinuity of the coverage of the correlation map, which is evident in the white areas representing missing information. The size of each patch is insufficient to capture significant dynamics, preventing the correlation matrix from encompassing enough information to reveal meaningful structures.  These artefacts become the most significant contributors to the variance and tend to appear as modes, although they have no physical meaning.
 
 When the patches of size $\mathcal{P} = D$ overlap by $50\%$ ($3^{rd}$ column) the reconstructed modes have a fair resemblance to the reference distribution. However, the first two modes suffer from artefacts at the edges of the overlapping regions. Modes $3$ and $4$ are reconstructed with good fidelity, while larger-scale modes ($5$ and $6$) partially capture smaller dynamic features but with noticeable distortion. Finally, when the overlap fraction is increased to $75\%$ ($4^{th}$ column) one observes that the discontinuous regions in modes $1$ and $2$ are attenuated. Yet the large-scale modes remain distorted, which is ascribed to the limited extent of the patch not being able to capture the mode distribution. In terms of fluid-dynamic understanding it is noteworthy that in the case of $75\%$ overlap modes $5$ and $6$ offer a clearer representation of vortex pairing dynamics, showing a structure analogous to that of mode $3$ and $4$ but with doubled wavelength. Remarkably, the local and truncated nature of the spatial correlation matrix might allow for a more local analysis reducing mode-mixing effects.
 
 The second set of patched reconstructions (Fig.~\ref{fig:phiJet_60}) deals with a larger patch size ($\mathcal{P} = 2D$). Here, the case with no overlap yields results comparable to those obtained with the smaller patch, with slight improvements. The larger patch size enables the capture of a more extensive portion of the domain, though it still restricts the representation of larger scales. It becomes evident that at such scale and with an overlap fraction of $75\%$ ($4^{th}$ column ) all the considered modes encompassing a larger area can be reconstructed at high fidelity. The exception being a cross-mode interference for the reconstruction of mode $2$ and $3$, where the high-wavenumber fluctuations due to shear layer instability are already appearing in these modes. Instead, for the reference distribution a more distinct separation is made between mode $1$-$2$ and modes $3$-$4$. This effect arises from the energy leakage between modes. The finite size of the patch inevitably limits its ability to capture large-scale structures entirely, resulting in an incomplete representation of the phenomenon. This impacts the variance induced by the phenomenon in the dataset, leading to an inaccurate eigenvalue distribution that affects the entire decomposition process. This effect is analogous to the spectral leakage phenomenon in Fourier analysis in finite-duration signals. Establishing a general guideline to mitigate cross-mode interference is challenging, as its impact is highly case-dependent. Larger patch sizes are more affected by this issue. When increasing the patch size is not feasible, enhancing the overlap between patches can also help in achieving smoother spatial modes.

From an overall perspective, the primary observation that arises from the analysis of Figs.~\ref{fig:phiJet_30}-\ref{fig:phiJet_60}, is that structures outside of the patch  are not visible when there is no overlap among adjacent patches. In particular, in absence of overlap, the portion of the observed correlation map for each point can be strongly skewed, especially near the edges of the patch. This results in modes with spatial discontinuities. However, even with fixed patch size, introducing some overlap reduces this effect and allows capturing structures covering more than a single patch.

This general behaviour is something already predictable from the analysis of $\mathbf{\hat{C}}$ in 
Fig.~\ref{fig:CorrelationJet}: the most relevant contributions to the correlation are located nearby the diagonal of the matrix. The comparison between Fig.~\ref{fig:CorrelationJet}.a and Fig.~\ref{fig:CorrelationJet}.g shows how this combination of $\mathcal{P}$ and $\mathcal{OF}$ is able to capture the highest portion and consequently the most energetic contribution of the correlation matrix. 

The energy distribution (Fig.~\ref{fig:eigjet}) also provides hints on the effect of partial measurement of the correlation map. The reference is shown as a continuous black line, while the two values of $\mathcal{P}$ under investigation are colour-coded in shades of blue ($\mathcal{P} = D$) and red ($\mathcal{P} = 2D$). Different line styles represent the levels of overlap: dashed for no overlap, dotted for $50\%$, and dash-dotted for $75\%$. In addition, empty colour-coded circles display the first six modes analysed in the previous comparison. The energy level associated with the most energetic modes is consistently lower than in the reference case for all tested conditions. Increasing patch size and overlap unveils a larger portion of the correlation matrix, thus producing energy distribution among modes closer to the full field measurement. It must be remarked that higher-order modes tend to experience instead the opposite effect, with a slight increase of energy for increasing patch size and overlap factor. This is likely due to noise, induced by discontinuities in the correlation map. 
Interestingly, the cause of the previously discussed cross-mode interference is particularly evident also here. Looking at the red curves ($\mathcal{P} = 2D$), the distance between the second and third eigenvalues is smaller compared to the reference, and it continues to decrease as the overlap is reduced. This diminished separation leads to a mixing of the two spatial modes.

\begin{figure}[t]
    \centering
    \begin{overpic}[width=0.7\textwidth, unit = 1mm]{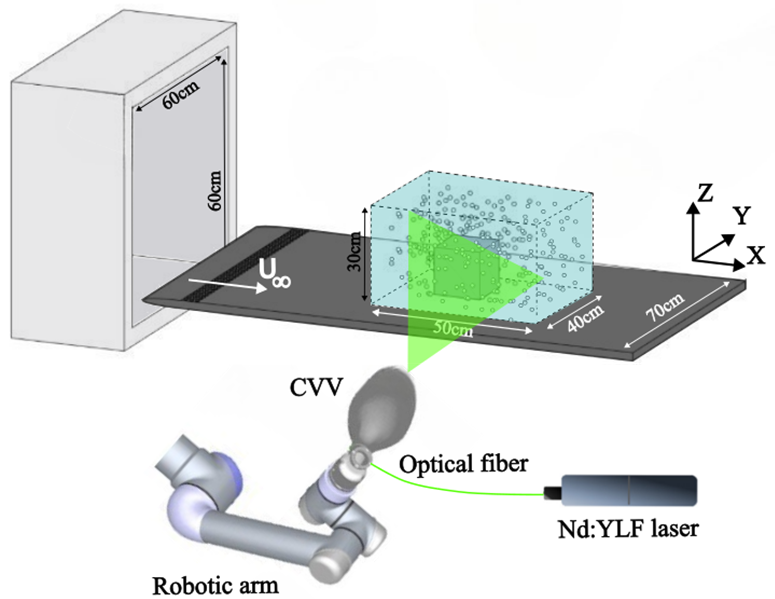}
    \put(0,70){\parbox{90mm}{\centering \textbf{{Robotic PIV}}}} 
    \end{overpic}
    \vskip 10mm
    \begin{overpic}[width=0.8\textwidth, unit = 1mm]{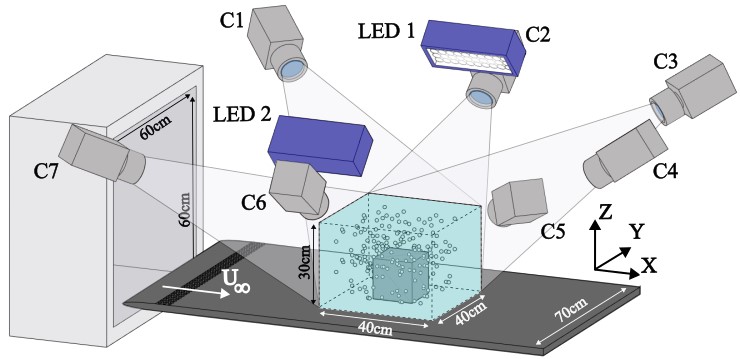}
    \put(0,54){\parbox{100mm}{\centering \textbf{{Large-aperture 3D-LPT }}}}
    \end{overpic}
    \caption{Sketch of the two experimental setups: on top the robotic PIV, on bottom the large-aperture 3D-LPT \citep{hendriksen2024object} used here as a reference.}
    \label{fig: setup}
\end{figure}

\begin{figure}[t]
    \centering
    \includegraphics[scale = 1]{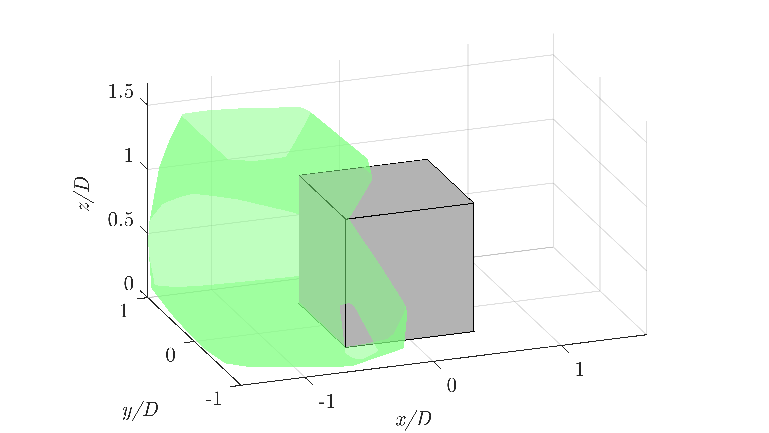}
    \caption{Example of robotic PIV measurement domain for a single patch visualized with the convex hull of the ensemble of tracked particles, $\mathcal{P}\approx1.6D$}
    \label{fig:patch}
\end{figure}

\begin{figure}[t]
    \centering
    \begin{overpic}[scale = 1,unit = 1mm]{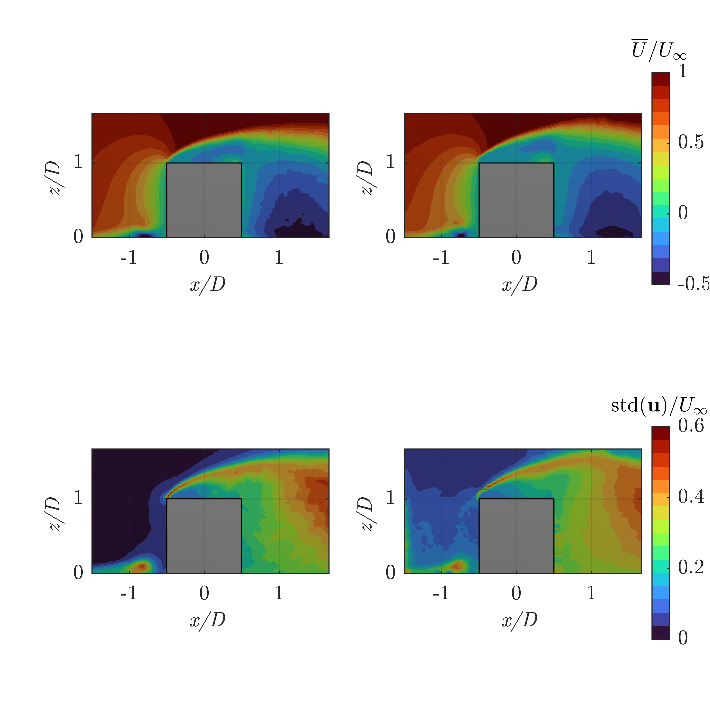}
    \put(10,116){\parbox{50mm}{\centering \textbf{{3D LPT}}}} 
     \put(65,116){\parbox{50mm}{\centering \textbf{{Robotic PIV }}}}
    \end{overpic}
    
    \caption{Top: ensemble average streamwise velocity  component using a bin size of  $b = 10$ mm, with uniform weight within the bin; bottom: standard deviation of the velocity magnitude. Left: reference result; right: robotic PIV result. Streamwise velocity and standard deviation are both normalized with the free-stream velocity  $ U_\infty$ and refer to the mid plane at $y/D = 0$.}
    \label{fig:mean_stat}
\end{figure}

\section{Application to 3D robotic PIV}
\label{3D}

\begin{figure}[t]
    \centering
    \includegraphics[scale = 1]{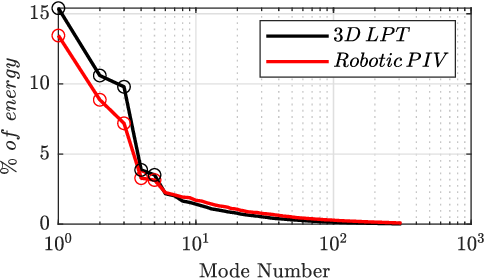}
    \caption{Eigenvalue distribution for Robotic PIV (red continuos line) and large-aperture 3D-LPT (black continuous line) normalized with the total amount of energy of the reference case (large-aperture 3D-LPT). Empty circles highlight the first $6$ modes.}
    \label{fig:eigCube}
\end{figure}

\begin{figure}[t]
    \centering
    \begin{overpic}[scale =1 , unit = 1mm]{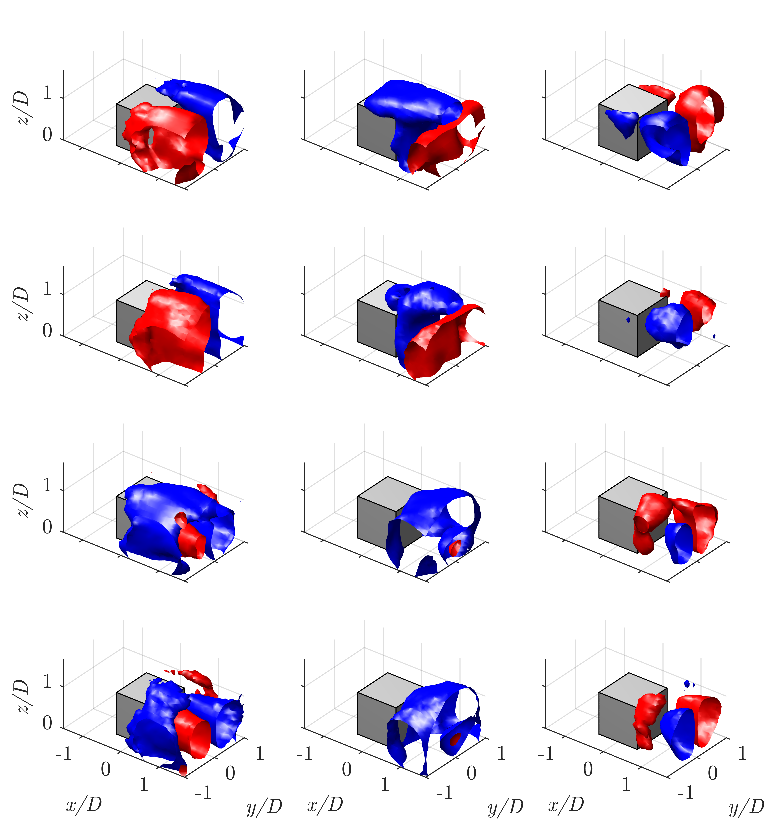} 

    \put(0,102){\parbox{18mm}{\centering \textbf{{Robotic PIV}}}} 
     \put(0,135){\parbox{18mm}{\centering \textbf{{3D LPT }}}}
     \put(10,135){\parbox{30mm}{\centering $\phi_{1u}$}}
     \put(50,135){\parbox{30mm}{\centering $\phi_{1v}$}}
     \put(90,135){\parbox{30mm}{\centering $\phi_{1w}$}}
     \put(10,103){\parbox{30mm}{\centering $\hat{\phi}_{1u}$}}
     \put(50,103){\parbox{30mm}{\centering $\hat{\phi}_{1v}$}}
     \put(90,103){\parbox{30mm}{\centering $\hat{\phi}_{1w}$}}

    \put(0,37.5){\parbox{18mm}{\centering \textbf{{Robotic PIV}}}} 
     \put(0,70){\parbox{18mm}{\centering \textbf{{3D LPT }}}}
     \put(10,70){\parbox{30mm}{\centering $\phi_{2u}$}}
     \put(50,70){\parbox{30mm}{\centering $\phi_{2v}$}}
     \put(90,70){\parbox{30mm}{\centering $\phi_{2w}$}}
     \put(10,38.5){\parbox{30mm}{\centering $\hat{\phi}_{2u}$}}
     \put(50,38.5){\parbox{30mm}{\centering $\hat{\phi}_{2v}$}}
     \put(90,38.5){\parbox{30mm}{\centering $\hat{\phi}_{2w}$}}
    \end{overpic}
   
    \caption{
    Isosurfaces of positive and negative velocity for each of the three components of the spatial POD modes $1$ and $2$. From the first row to the last one: mode $1$, $3D$ LPT; mode $1$, Robotic PIV; mode $2$, $3D$ LPT;  mode $2$, Robotic PIV.}

    \label{fig:Q1}
\end{figure}

\begin{figure}[t]
    \centering
    \begin{overpic}[scale =1 , unit = 1mm]{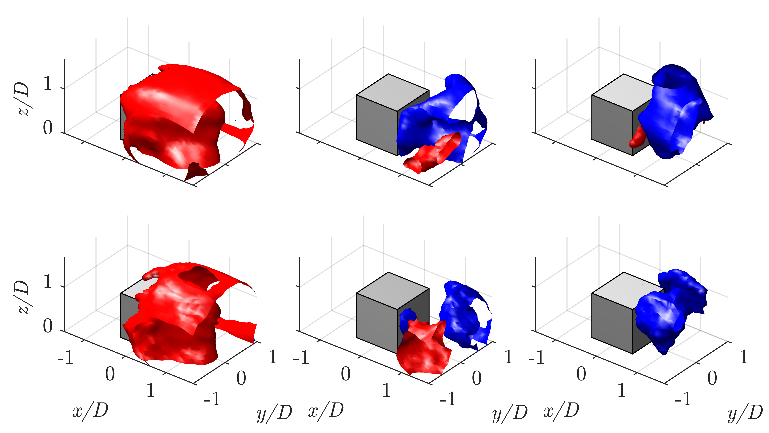}

     \put(0,36.5){\parbox{18mm}{\centering \textbf{{Robotic PIV}}}} 
     \put(0,70){\parbox{18mm}{\centering \textbf{{3D LPT }}}}
     \put(10,70){\parbox{30mm}{\centering $\phi_{3u}$}}
     \put(50,70){\parbox{30mm}{\centering $\phi_{3v}$}}
     \put(90,70){\parbox{30mm}{\centering $\phi_{3w}$}}
     \put(10,37.5){\parbox{30mm}{\centering $\hat{\phi}_{3u}$}}
     \put(50,37.5){\parbox{30mm}{\centering $\hat{\phi}_{3v}$}}
     \put(90,37.5){\parbox{30mm}{\centering $\hat{\phi}_{3w}$}}
    \end{overpic}
    \caption{Isosurfaces of positive and negative velocity for each of the three components of the spatial POD mode $3$. Top: $3D$ LPT; bottom: Robotic PIV.} 
    \label{fig:Q23}
\end{figure}

\begin{figure}[t]
    \centering
    \begin{overpic}[scale =1 , unit = 1mm]{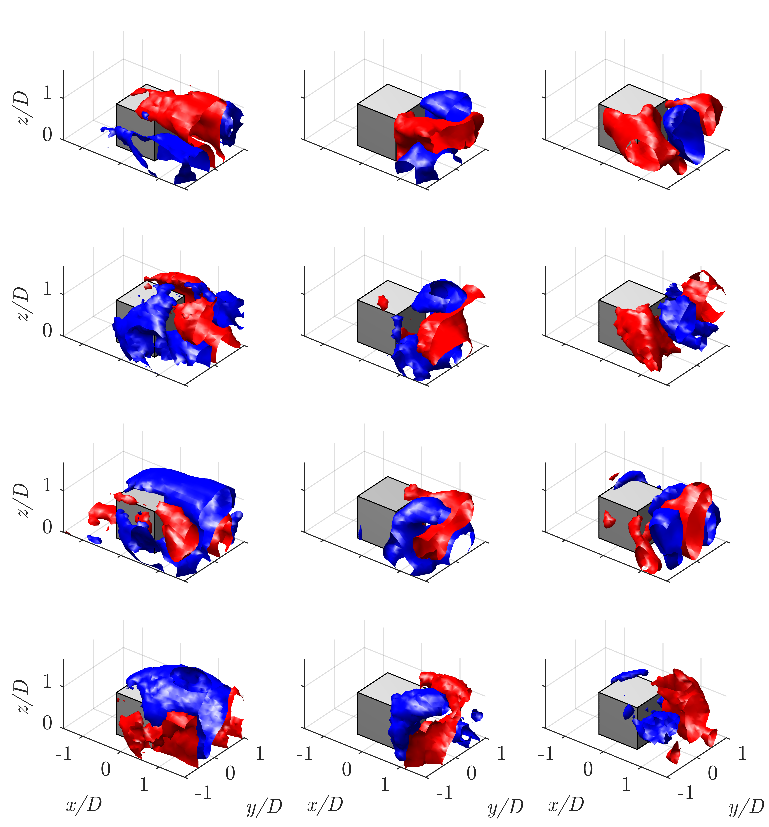} 
     \put(0,102){\parbox{18mm}{\centering \textbf{{Robotic PIV}}}} 
     \put(0,135){\parbox{18mm}{\centering \textbf{{3D LPT }}}}
     \put(10,135){\parbox{30mm}{\centering $\phi_{4u}$}}
     \put(50,135){\parbox{30mm}{\centering $\phi_{4v}$}}
     \put(90,135){\parbox{30mm}{\centering $\phi_{4w}$}}
     \put(10,103){\parbox{30mm}{\centering $\hat{\phi}_{4u}$}}
     \put(50,103){\parbox{30mm}{\centering $\hat{\phi}_{4v}$}}
     \put(90,103){\parbox{30mm}{\centering $\hat{\phi}_{4w}$}}

     \put(0,36.5){\parbox{18mm}{\centering \textbf{{Robotic PIV}}}} 
     \put(0,70){\parbox{18mm}{\centering \textbf{{3D LPT }}}}
     \put(10,70){\parbox{30mm}{\centering $\phi_{5u}$}}
     \put(50,70){\parbox{30mm}{\centering $\phi_{5v}$}}
     \put(90,70){\parbox{30mm}{\centering $\phi_{5w}$}}
     \put(10,37.5){\parbox{30mm}{\centering $\hat{\phi}_{5u}$}}
     \put(50,37.5){\parbox{30mm}{\centering $\hat{\phi}_{5v}$}}
     \put(90,37.5){\parbox{30mm}{\centering $\hat{\phi}_{5w}$}}
    \end{overpic}
    \caption{Isosurfaces of positive and negative velocity for each of the three components of the spatial POD modes $4$ and $5$. From the first row to the last one: mode $4$, $3D$ LPT; mode $4$, Robotic PIV; mode $5$, $3D$ LPT;  mode $5$, Robotic PIV. }
    \label{fig:Q45}
\end{figure}

As introduced above, three-dimensional measurements suffer often from a limited measurement domain as a result of limited illumination power, imaging resolution of volumetric particle fields. Robotic volumetric PIV \citep{jux2018robotic} covers larger volumes by partitioning the domain into smaller sub-regions based on rapid, robotic manipulation of the optical head (coaxial velocimeter, \citet{schneiders2018coaxial}). It is here investigated whether robotic PIV measurements can be connected for modal reconstruction extending the patched POD technique to three-dimensional data. The experimental validation compares two experiments conducted to measure the three-dimensional flow field around a wall-mounted cube of side length $120$ mm. Reference data is obtained with Lagrangian particle tracking measurements performed with $7$ cameras imaging simultaneously the full domain \citep{hendriksen2024object}. The same flow field is surveyed with Robotic PIV covering the domain with a multitude of views each covering approximately $1.5$ cube side lengths. 

Both experimental campaigns were conducted at the TU Delft Aerodynamics Laboratories in an open-jet, open-circuit, low-speed wind tunnel (W-tunnel). The freestream velocity is set at $10$ m/s, which corresponds to $Re = 81,000$ based on the cube side.

HFSBs were used as tracer particles \citep{bosbach2009large}. The HFSBs seeding generator comprises $10$ parallel wings, each $1$ meter in span and equipped with $20$ nozzles spaced $5$ cm apart. This setup results in a seeding rake with a total of $204$ nozzles, covering a seeding surface area of approximately $0.50 \times 0.25 m^2$ in the wind tunnel's settling chamber. The generator, operated by a in-house fluid supply unit, controls the flow rate of air, helium, and soap.
The nominal diameter of the tracer particles is $500 \mu m$.

\begin{table}[t]
\centering
\caption{Experimental setup details.}
\begin{tabularx}{\textwidth}{|c|X|X|}
\hline
{} & \centering\textbf{Robotic PIV} & \textbf{\,\,\,\,\,\,\,\,\,\,\,\,\,\,\,\,\,\,\,\,\,\,\,\,\,\,\,\,\,\,3D LPT}\\ \hline
\textbf{Seeding} & HFSB, $d = 500\,\mu$\text{m} & HFSB, $d = 500\,\mu$\text{m} \\ \hline
\textbf{Working fluid} & Air & Air\\ \hline
\textbf{Characteristic
length} & $D = 120\,mm$ & $D = 120\,mm$ \\ \hline
\textbf{Exit velocity} & $U_\infty = 10\, \text{m/s}$ &$U_\infty = 10\, \text{m/s}$ \\ \hline
\textbf{Reynolds number} & $Re_D = 81,000$ & $Re_D = 81,000$ \\ \hline
\textbf{Illumination} & Quantronix Darwin Duo Nd:YLF diode-pumped laser ($\lambda$ = $527$ nm, $2 \times 25$ mJ pulse energy at
$1$ kHz) & $2 \times$ LaVision flashlight-$300$ LEDs ($33\,\mu\text{s}$ pulse duration)\\ \hline
\textbf{Imaging} & $4\times$ CMOS cameras, objective of $4$ mm focal length, integrated within the LaVision MiniShaker Aero & $7\times$ CMOS cameras ($1$\,Mpx, focal length $4\times 60\,mm$ and $3\times 50\,mm$) \\ \hline
\textbf{Processing} & 2P-STB with the aid of a predictor \citep{saredi2020multi} & TR-STB \citep{schanz2016shake}\\ \hline
\end{tabularx}
\label{tab:cube}
\end{table}

The two experimental setups are sketched in Fig.~\ref{fig: setup} and summarized in Tab.~\ref{tab:cube}. 
The robotic system velocimeter probe consists of four CMOS cameras equipped with an objective of $4$ mm focal length, integrated within the LaVision MiniShaker Aero. The illumination of the investigated domain is provided by a Quantronix Darwin Duo Nd:YLF diode-pumped laser ($\lambda$ = $527$ nm, $2 \times 25$ mJ pulse energy at
$1$ kHz). Measurements in $35$ patches with high overlap ($\approx70-80\%$ ) have been carried out. For each patch, $5000$ image pairs are captured in frame-straddling mode, at an acquisition frequency of $200$ Hz. The patches cover an investigation region spanning $x \in [-180,300]$  mm, $y \in[-180,180]$ mm and $z \in[0,240]$ mm. As shown in Fig.~\ref{fig:patch}, the size of each patch is approximately $200$ mm ($\approx1.6D$). Ensuring identical overlap among all the patches is not feasible, as the robotic system requires manual selection and storage of the position of each cone to build the investigation sequence and then compute the optimal path to achieve it. To give the reader an estimation of the overlap factor, the following metric has been employed to compute the overlap among two consecutive patches $i$ and $j$:

\begin{equation}
    \mathcal{OF}_{ij} = \frac{\left(\mathcal{V}_i\cap\mathcal{V}_j\right)}{0.5\left(\mathcal{V}_i+\mathcal{V}_j\right)},
\end{equation}

where $\mathcal{V}$ denotes the volume of the patch. Averaging the values across all the patches in the sequence leads to $\mathcal{OF} \approx 75\%$.
For further details about the robotic system, the reader is referred to \citet{jux2018robotic}. The acquired images are then analyzed using the Shake-The-Box algorithm for double-frame recordings, using a velocity predictor \citep{saredi2020multi}.

The experimental setup of the large-aperture 3D-LPT experiments used as reference is detailed in the work of \citet{hendriksen2024object}. For this measurement, $7$ high-speed CMOS cameras were used while the illumination was provided by two pulsed LEDs. A total amount of $10,000$ snapshots (two independent sequences of $5000$ snapshots each) were acquired at a frequency of $3$ kHz. The investigated domain in this case has the same spanwise and wall-normal extension as in the robotic-PIV experiment, but a shorter streamwise extension ($x\ \in[-180,\ 200]$ mm).
To perform a fair comparison between the results of the two measurements, all the instantaneous data are binned onto the same grid ($x/D\ \in[-1.5D,1.67D]$ , $y/D\ \in[-1D,1D]$ and $z/D\ \in[0D,1.67D]$ ) with a bin size of $D/3$ (or $40$ mm) and $75\%$ overlapping. 

In Fig.~\ref{fig:mean_stat} the mean flow distribution in the mid-plane is shown for the two measurements. The results are obtained by ensemble averaging the particle velocities within cells (cubic bins) of $10$ mm side length. The mean flow does not show any appreciable difference between robotic PIV and reference $3D$-LPT measurements, indicating a high degree of repeatability and statistical convergence. The flow exhibits a forward stagnation point at the cube front face and then  accelerates along said face to separate from the sharp leading edge. The resulting shear layer does not reattach on the upper face, but rather downstream of the cube. Such a reattachment line is not captured within the domain of measurement. The overall topology is rather complex with a horseshoe vortex dominating the wall region surrounding the cube and an upper separated region that encloses an arch-like reverse-flow structure. Details of this flow have been studied and reported in the literature and the reader is referred to \citet{martinuzzi1993flow} and \citet{schroder2020flow} among others.

Figure~\ref{fig:mean_stat}-bottom-left illustrates the statistical fluctuations of the streamwise velocity component measured from 3D-LPT. The fluctuations are concentrated in the near-wall region of the incoming turbulent boundary layer with a local peak at the inception of the horseshoe vortex, indicating its unstable position. Furthermore, significantly higher levels of fluctuations are encountered in the separated shear layer (approx. $40\%$) and in the trailing separated region (exceeding $50\%$). In this case, a marked difference is observed in the outer stream, where the robotic PIV measurements (Fig.~\ref{fig:mean_stat}-bottom-right) yield a higher level of fluctuations (approximately $5\%$) ascribed to measurement noise. Besides such differences, the overall distribution of velocity fluctuations exhibits a sufficiently well-repeated pattern and we expect that the impact of these differences on the modal decomposition of the large-scale fluctuations should be minimal, and presumably limited to the regions ahead of the cube. It should be retained in mind that in this case the 3D-LPT measurements benefit from the large-aperture imaging (viz. triangulation) of the particle tracers and the time-resolved analysis of Shake the Box, yielding significantly augmented accuracy to the measurement compared to the robotic system employing a coaxial velocimeter \citep{schneiders2018coaxial}. Thus the lower level of free stream turbulence (expected below $2\%$) is captured by 3D-LPT and not with the robotic PIV measurement.


Figure \ref{fig:eigCube} presents the eigenvalue distributions for both cases, normalized with the total amount of energy from the 3D-LPT measurements case. The figure illustrates that the Robotic PIV yields a lower energy level in the first modes. As discussed in the previous section, this is mostly ascribed to the truncation of the spatial correlation due to the finite size of the patch ($\approx 37\%$ of the total volume). 

A comparison among the first $5$ spatial modes in terms of velocity components is performed to assess the results.

As reported in the study by \citet{da2024revisiting}, common features of all surface-mounted finite-height square prisms include the horseshoe vortex and the arch vortex, followed by dipole/quadrupole vortical structures in the wake. Their position, strength, and behaviour are primarily dependent on the aspect ratio ($AR$) of the prism, boundary layer thickness ($\delta/D$), and the Reynolds number investigated.
For the specific case of the cube, which can be considered a prism with $AR = 1$, the main structure is the arch vortex \citep{martinuzzi1993flow}. This structure is easily recognizable by the presence of two legs, where the flow rotates around wall-normal (vertical) axes, and a roof, where the rotation is around a horizontal axis perpendicular to the free-stream direction. The experimental study of this vortex was carried out by \citet{becker2002flow}, while \citet{kawai2012near} captured the temporal evolution of this vortex using stereoscopic PIV, revealing characteristics similar to the von Kármán street.

 Figure \ref{fig:Q1} presents the first two spatial modes, collecting $20-25\%$ of the total energy. The coherent structure represented by these modes primarily captures a dominant vortex-shedding mechanism, as also reported by \citet{da2024flow}. This vortex shedding is characterized by the formation of an arch vortex due to the interaction between the separated shear layers and the wake region behind the cube, as mentioned before, resembling a von Kármán street.

Figure \ref{fig:Q23} illustrates the third spatial mode, which, consistent with the spectral analysis by \citet{da2024flow}, exhibits characteristics of a drift mode. This mode accounts for approximately $7-10\%$ of the total energy and plays a key role in the temporal ``drifting'' behaviour of the near wake. According to \citet{wang2021experimental}, this mode is primarily responsible for the periodic elongation and contraction of the near wake. These findings are in agreement with the observations of \citet{bourgeois2013generalized}, further validating the influence of the mode on wake dynamics.

For a deeper discussion about the role of these structures in the flow dynamics around the wall-mounted cube, the reader is referred to \citet{da2024flow}.

The interpretation of modes $4$ and $5$ in Fig.~\ref{fig:Q45} is more challenging, as they are likely associated with a secondary shedding pattern occurring at lower frequencies. However, a detailed analysis of the flow behavior is beyond the scope of this work. Instead, the focus is on ensuring that the modes obtained from locally asynchronous patches of Robotic PIV acquisition align with those derived from the 3D LPT measurements over the entire region of interest. In this context, the modes show a strong agreement, extending to higher-order modes not presented here for brevity. 

\section{Conclusions}
A methodology to achieve full-domain POD spatial modes starting from asynchronous measurements in different flow regions of a domain is proposed. The technique is based on unveiling the bulk of the content of the two-point correlation matrix. Introducing overlap between patches has been shown to improve the estimation of the mode energy, reduce the unphysical discontinuities between regions in the global modes, and increase the range of reconstructed scales even beyond the patch size. The approach, named Patch POD, is targeted explicitly to estimation of large-scale modes. 

A validation has been carried out by dissecting $2D$ measurements of a submerged jet in patches. The POD modes obtained with our method using solely the patches match reasonably well with the results of a modal decomposition run on the entire domain. Cross-mode interference could appear as a drawback due to the limited size of the patch. However, increasing the patch size, when feasible, or enhancing the overlap between patches can help mitigate this effect. An application to a volumetric robotic PIV experiment on the flow around a wall-mounted cube has been also carried out. The results show a good agreement between the modes identified by merging asynchronous patches and those obtained by a large-aperture 3D-LPT covering the entire domain. This approach demonstrates the ability to recover spatial structures that are almost the double of the patch size, and opens the possibility of obtaining global modes on a larger scale than the volume observed by each realization.

\section*{Acknowledgment}
This project has received funding from the European Research Council (ERC) under the European Union’s Horizon 2020 research and innovation program (grant agreement No 949085). Views and opinions expressed are however those of the authors only and do not necessarily reflect those of the European Union or the European Research Council. Neither the European Union nor the granting authority can be held responsible for them. Luca Franceschelli and Qihong Lorena Li Hu are kindly acknowledged for providing the experimental dataset of the jet flow. Luuk Hendriksen is kindly acknowledged for making available the 3D-LPT dataset of the wall mounted cube.

\bibliography{sn-bibliography}

\end{document}